\documentclass[twoside,twocolumn,9pt]{article}
\usepackage{extsizes}
\usepackage[super,sort&compress,comma]{natbib} 
\usepackage[version=3]{mhchem}
\usepackage[left=1.5cm, right=1.5cm, top=1.785cm, bottom=2.0cm]{geometry}
\usepackage{balance}
\usepackage{times,mathptmx}
\usepackage{sectsty}
\usepackage{graphicx} 
\usepackage{lastpage}
\usepackage[format=plain,justification=justified,singlelinecheck=false,font={stretch=1.125,small,sf},labelfont=bf,labelsep=space]{caption}
\usepackage{float}
\usepackage{fancyhdr}
\usepackage{fnpos}
\usepackage[english]{babel}
\addto{\captionsenglish}{%
  
}
\usepackage{array}
\usepackage{droidsans}
\usepackage{charter}
\usepackage[T1]{fontenc}
\usepackage[usenames,dvipsnames]{xcolor}
\usepackage{setspace}
\usepackage[compact]{titlesec}
\usepackage[colorlinks]{hyperref}
\usepackage{siunitx} 

\usepackage{epstopdf}

\definecolor{cream}{RGB}{222,217,201}

\begin{document}

\pagestyle{fancy}
\thispagestyle{plain}
\fancypagestyle{plain}{

\renewcommand{\headrulewidth}{0pt}
}

\makeFNbottom
\makeatletter
\renewcommand\LARGE{\@setfontsize\LARGE{15pt}{17}}
\renewcommand\Large{\@setfontsize\Large{12pt}{14}}
\renewcommand\large{\@setfontsize\large{10pt}{12}}
\renewcommand\footnotesize{\@setfontsize\footnotesize{7pt}{10}}
\makeatother

\renewcommand{\thefootnote}{\fnsymbol{footnote}}
\renewcommand\footnoterule{\vspace*{1pt}%
\color{cream}\hrule width 3.5in height 0.4pt \color{black}\vspace*{5pt}} 
\setcounter{secnumdepth}{5}

\makeatletter 
\renewcommand\@biblabel[1]{#1}            
\renewcommand\@makefntext[1]%
{\noindent\makebox[0pt][r]{\@thefnmark\,}#1}
\makeatother 
\renewcommand{\figurename}{\small{Fig.}~}
\sectionfont{\sffamily\Large}
\subsectionfont{\normalsize}
\subsubsectionfont{\bf}
\setstretch{1.125} 
\setlength{\skip\footins}{0.8cm}
\setlength{\footnotesep}{0.25cm}
\setlength{\jot}{10pt}
\titlespacing*{\section}{0pt}{4pt}{4pt}
\titlespacing*{\subsection}{0pt}{15pt}{1pt}

\fancyfoot{}
\fancyfoot[LO,RE]{\vspace{-7.1pt}\includegraphics[height=9pt]{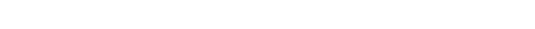}}
\fancyfoot[CO]{\vspace{-7.1pt}\hspace{13.2cm}\includegraphics{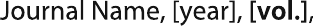}}
\fancyfoot[CE]{\vspace{-7.2pt}\hspace{-14.2cm}\includegraphics{RF}}
\fancyfoot[RO]{\footnotesize{\sffamily{1--\pageref{LastPage} ~\textbar  \hspace{2pt}\thepage}}}
\fancyfoot[LE]{\footnotesize{\sffamily{\thepage~\textbar\hspace{3.45cm} 1--\pageref{LastPage}}}}
\fancyhead{}
\renewcommand{\headrulewidth}{0pt} 
\renewcommand{\footrulewidth}{0pt}
\setlength{\arrayrulewidth}{1pt}
\setlength{\columnsep}{6.5mm}
\setlength\bibsep{1pt}

\makeatletter 
\newlength{\figrulesep} 
\setlength{\figrulesep}{0.5\textfloatsep} 

\newcommand{\topfigrule}{\vspace*{-1pt}%
\noindent{\color{cream}\rule[-\figrulesep]{\columnwidth}{1.5pt}} }

\newcommand{\botfigrule}{\vspace*{-2pt}%
\noindent{\color{cream}\rule[\figrulesep]{\columnwidth}{1.5pt}} }

\newcommand{\dblfigrule}{\vspace*{-1pt}%
\noindent{\color{cream}\rule[-\figrulesep]{\textwidth}{1.5pt}} }

\makeatother

\twocolumn[
  \begin{@twocolumnfalse}
    {\includegraphics[height=30pt]{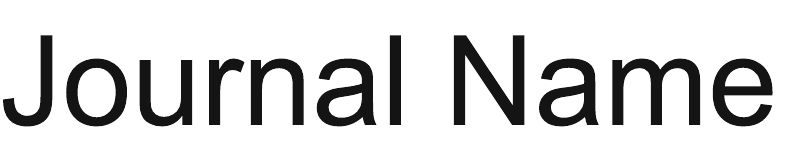}\hfill%
 \raisebox{0pt}[0pt][0pt]{\includegraphics[height=55pt]{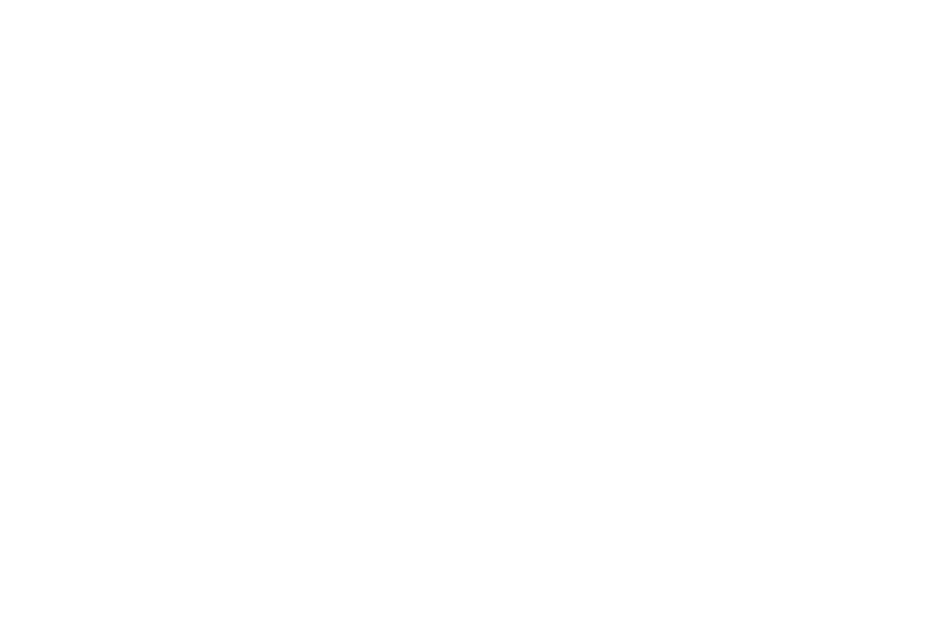}}%
 \\[1ex]%
 \includegraphics[width=18.5cm]{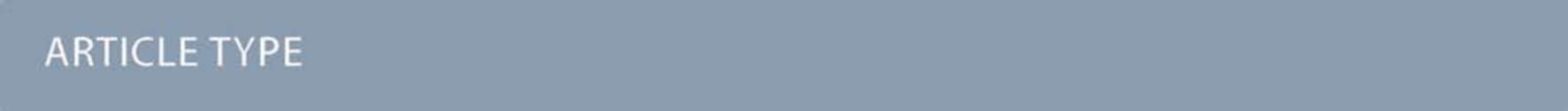}}\par
\vspace{1em}
\sffamily
\begin{tabular}{m{4.5cm} p{13.5cm} }

\includegraphics{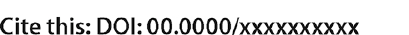} & \noindent\LARGE{\textbf{Optimal quantum dot size for photovoltaics with fusion}} \\
\vspace{0.3cm} & \vspace{0.3cm} \\

& \noindent\large{Benedicta Sherrie,\textit{$^{a}$} Alison M. Funston,\textit{$^{a}$} and Laszlo Frazer$^{\ast}$\textit{$^{a}$}} \\

\includegraphics{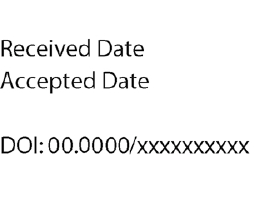} & \noindent\normalsize{
	Light fusion increases the efficiency of solar cells by converting photons with lower energy than the bandgap into higher energy photons.  The solar cell converts the product photons to current.  We use Monte Carlo simulation to predict that lead sulfide  (\ce{PbS}) quantum dot sensitizers will enable fusion with a figure of merit on the \si{\milli\ampere\per\square\cm} scale, exceeding current records, while enabling silicon cell compatibility.  Performance is highly sensitive to quantum dot size, on the order of \si{\milli\ampere\per\square\cm\per\nm}.

	\href{https://doi.org/10.1039/D0NR07061K}{Sherrie, B., Funston, A. M., \& Frazer, L. (2020). Optimal quantum dot size for photovoltaics with fusion. Nanoscale, 12(48), 24362-24367.}
} \\

\end{tabular}

 \end{@twocolumnfalse} \vspace{0.6cm}

  ]

\renewcommand*\rmdefault{bch}\normalfont\upshape
\rmfamily
\section*{}
\vspace{-1cm}


\footnotetext{\textit{$^{a}$~ARC Centre of Excellence in Exciton Science and School of Chemistry, Monash University, 17 Rainforest Walk, Clayton, Vic 3800, Australia. Tel: 61 401 648 058; E-mail: laszlo.frazer@monash.edu, laszlo@laszlofrazer.com}}




\section{Introduction}

Solar cells recently became the cheapest source of electricity.\cite{lazard}  Increasing solar cell efficiency will decrease energy costs.\cite{peters2019value}  For wider bandgap solar cells, the main inefficiency is transparency.\cite{tayebjee2015beyond,shockley1961detailed}  Here, we simulate the use of absorptive and tunable lead sulfide quantum dots to capture and convert the wasted light.

\begin{figure}[h]
\centering
\includegraphics[width=.8\linewidth]{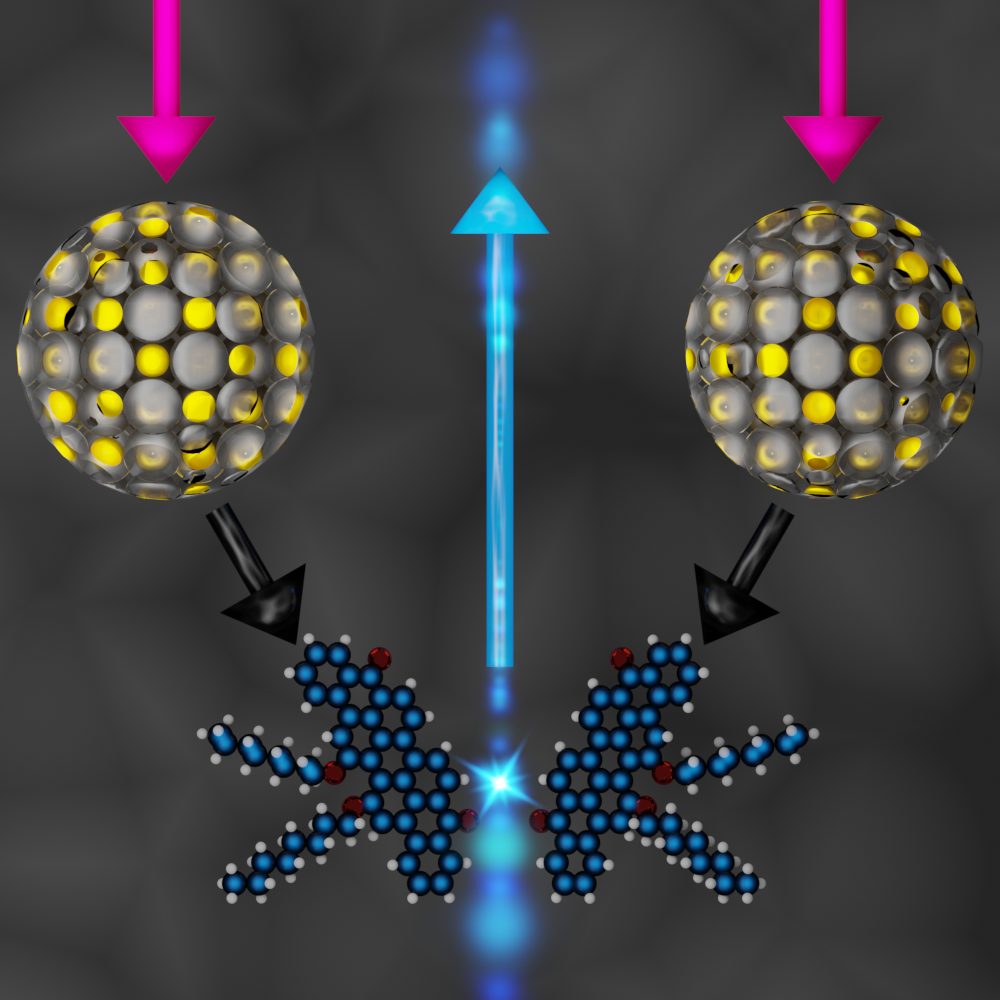}
\caption{Schematic illustration of the device architecture.  Sunlight illuminates the device from the top (red arrows).  A solar cell (not shown), which is transparent below its bandgap, filters the sunlight.  Beneath the solar cell, a quantum dot sensitizer (halite structured spheres) captures the light.  Quanta of energy are transferred (black arrows) to a hypothetical light fusion emitter (molecular model) which  converts light from below the solar cell bandgap to above the solar cell bandgap.  The upconverted light travels up to the solar cell (blue arrow).  The sensitizer and emitter are uniformly distributed.}
\label{fig:diagram}
\end{figure}

Triplet fusion,\cite{ravetz2019photoredox} also known as triplet-triplet annihilation upconversion, photochemical upconversion,\cite{dilbeck2018molecular,zeng2017molecular} or Auger annihilation, utilizes light that passes through a solar cell.  Triplet fusion uses two chemical species: a sensitizer and an emitter, as illustrated in Fig. \ref{fig:diagram}.  In a well-designed system, the sensitizer captures solar spectral irradiance which passes through the solar cell; \emph{i.e.} the photons which lie below the cell bandgap.  For sensitizers comprising semiconducting nanoparticles, the resulting excitons are transferred to the emitter triplet state.  The emitter fuses the triplet excitons, producing higher energy singlet excitons.  The fusion system fluoresces, releasing upconverted photons with an energy above the cell bandgap.  The net effect is that photons spontaneously increase in energy, enabling higher solar cell efficiency.\cite{tayebjee2015beyond}  Alternate approaches to the utilization of sub-bandgap photons include multijunction\cite{leijtens2018opportunities,werner2018perovskite} and intermediate band devices.\cite{okada2015intermediate}

Fusion requires the annihilation of excitons, which only happens when the excitons are close together.  To achieve a high density of excitons, it is essential to capture as much light as possible in a small volume.\cite{schmidt2014photochemical}  This can be achieved by use of a sensitizer with a high molar absorption coefficient and excitation rate.  Determining the excitation rate from the  absorption spectrum of the sensitizer and spectral irradiance of the sun is typically complicated, motivating our Monte Carlo simulations as a method of sensitizer selection.

Several groups have recently shown experimentally that quantum dots are superior to traditional organic sensitizers.\cite{nienhaus2017speed,huang2015hybrid,mase2017triplet,yanai2017new,huang2018semiconductor,nienhaus2018using,nishimura2019photon,okumura2019visible,he2019visible,de2020anthracene,beery2019cdse,gholizadehoxygen,huang2020evolution,lai2020red}  The very large absorption coefficient of quantum dots outweighs their tendency to reabsorb converted light.\cite{frazer2017optimizing} The absorption coefficient is not large enough to capture substantial light with one layer of particles.  Quantum dots have the additional advantages that they are highly photostable, have an absorption cross-section which is tunable across a wide range of energies and, in some cases, are already used in the electronics industry.

One can synthesize quantum dots from several semiconductor materials.  
Group IV quantum dots tend to have energy levels that are too high for our purposes.  III-V quantum dots tend to be difficult to synthesize on a large scale.  II-VI quantum dots are highly tunable across the visible spectrum, but cannot easily be tuned to absorb light below the bandgap of silicon, the most common solar cell absorber.  Perovskite quantum dots have a similar limitation, but are tuned with a different mechanism.  Cu-III-\ce{VI2} chalcopyrite and IV-VI quantum dots do not have these problems.  In chalcopyrites, the valence band maximum is at the $\Gamma$ point,\cite{soni2010electronic} while in IV-VI systems it is at the degenerate L point of the Brillouin zone.\cite{wei1997electronic}  The additional degeneracy and smaller unit cell of IV-VI valence electrons implies they will always have a higher molar absorption coefficient.  This probably cannot be overcome with differences in quantum dot density.  From the IV-VI family, we select elementally abundant and commonly synthesized \ce{PbS} to investigate.  \ce{PbSe} is very similar.\cite{moreels2009size,moreels2007composition}

\begin{figure}[h]
\centering
\includegraphics[width=\linewidth]{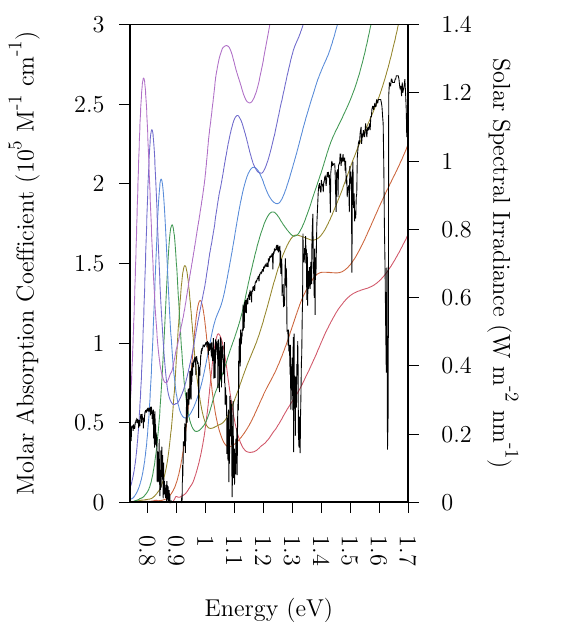}
\caption{Molar absorption coefficient of quantum dots with different radii\cite{wei1997electronic} (colors) and the AM1.5G solar spectral irradiance (black).  Bandgaps of interest for light fusion range from \SIrange{1}{1.5}{\electronvolt}.\cite{tayebjee2015beyond}  The major absorption lines displayed in the solar spectral irradiance are the overtones and combination bands of vibrational modes of water. The bandgap of silicon is 1.1 eV.}
\label{fig:absorption}
\end{figure}

Quantum confinement dictates the electronic structure of quantum dots.  Synthesis parameters control the size of the nanoparticles and therefore the energy levels of the quantum dot. One can synthesize PbS nanoparticles with a wide range of molar absorption coefficients (or absorption peak locations), as illustrated in Fig. \ref{fig:absorption}.  The molar absorption coefficient and peak locations cannot be controlled independently; both arise from chemical structure.  However, it is laborious to test the sensitization performance of a large number of quantum dot radii to find the ideal sensitizer.

While the overall efficiency of fusion systems involves an interplay between the properties of the sensitizer,\cite{gholizadeh2018photochemical} emitter,\cite{gray2015photophysical}, and environment,\cite{xu2018ratiometric,frazerphotochemical} here we focus on light capture by the sensitizer.  Previous work shows that light capture is essential.\cite{schmidt2014photochemical}  We quantitatively relate light capture to performance.

Here, we simulate fusion sensitized by lead sulfilde quantum dots.  We determine the optimal quantum dot size under the assumption that a hypothetical emitter with reasonable properties is used.  We consider how the performance of the fusion system depends on the density of the quantum dots and the bandgap of the solar cell.  We quantify device performance using the figure of merit,\cite{cheng2012improving} which is the photocurrent added by upconversion to a solar cell with perfect external quantum efficiency.  At solar irradiance, fusion is considered to be ``device relevant'' \cite{frazer2017optimizing} and economically beneficial \cite{peters2019value} at a figure of merit of \SI{0.1}{\milli\ampere\per\square\cm}.  The best experimental fusion figure of merit we are aware of is \SI{0.158}{\milli\ampere\per\square\cm},\cite{dilbeck2018molecular} just above this lower limit.  Thus far, the best devices used organic sensitizers incompatible with silicon cells.  They are incompatible because they do not absorb in the infrared beyond \SI{1100}{\nano\meter}. Our calculations predict that devices incorporating \ce{PbS} sensitizers will be able to achieve a new record for the figure of merit, while simultaneously meeting the requirements for silicon cell compatibility.

\section{Methods}
To determine optimal sensitizer parameters, we generated spectra for arbitrary quantum dot sizes. \ce{PbS} absorption spectra from Ref. \citenum{cademartiri2006size} were transformed to reduce the apparent size dependence. The energy scale was transformed to place the $E_1$ and $E_4$ peaks \cite{cademartiri2006size} at the same locations and the absorption scale was transformed so all $E_1$ peaks were the same amplitude.  Cubic radial basis function interpolation \cite{2020SciPy-NMeth} was used as a function of radius.  The generated spectra were transformed into molar absorption spectra using the energy level and extinction formulas in Ref. \citenum{cademartiri2006size}, restoring the size dependence.  Examples are shown in Fig. \ref{fig:absorption}.  The exact meaning of quantum dot radius is stoichiometry-dependent.\cite{moreels2009size}  We use the definition in Ref. \citenum{cademartiri2006size}.

The device figure of merit was computed using the previously reported algorithm.\cite{jefferies2019photochemical}   In brief, we used \num{e9} Monte Carlo samples of the AM1.5G \SI{1}{\kilo\watt\per\square\meter} solar spectrum (Other values are used in Supplementary Fig. 4.) with a sharp cutoff of \SI{1100}{\nano\meter} as a transparency model for a thick silicon solar cell.  The spatial distribution of absorption and self-absorption of quantum dots was computed at \num{e5} locations. No energy loss caused by the presence of trap states was included.  A reflector was included at the optimal location.\cite{frazer2017optimizing}  

Since this paper focusses on sensitization, we modeled hypothetical properties that a sensible emitter molecule would possess.  We assumed the emitter molecule would have no self-absorption; large sensitizer self-absorption makes emitter absorption negligible.  The emitter molecule emission was assumed to be evenly spread between 10 and 60 nm above the solar cell bandgap to approximate the observed insensitivity to absorption spectrum details. We assumed the emitter first order decay rate was \SI{e3}{\per\second} and the second order decay rate was \SI{4.7e-12}{\cm\cubed\per\second} (Ref. \citenum{frazer2017optimizing}) in imitation of diphenylanthracene.\cite{gray2017loss,gray2015photophysical} 
The results are generalized by calculating the performance of devices using a wide range of emitter first and second order decay rates.  This is displayed in Supplementary Figs. 2 and 3.
Energy transfer to the uniformly distributed emitter was assumed to be exothermic and in dynamic equilibrium.  The yield for production of fluorescence from triplet annihilation was assumed to be one.  While these properties have each been demonstrated in different emitters, they have not all been demonstrated in the same molecule.  Thermal and concentration quenching effects were omitted for lack of information.  The intensity of upconversion fluorescence was simulated according to the established rate equations.\cite{schmidt2014photochemical}

We assumed the concentration of quantum dots was limited by the Kepler theorem for close packed spheres. To leave room for passivating ligands, the radius associated with each particle was optimistically assumed to be \SI{0.5}{\nano\meter} larger than the quantum dot radius.  Close packing leaves 26\% of the volume available for exciton transport and emitter materials.

\section{Results and Discussion}

\begin{figure}[h]
\centering
\includegraphics[width=\linewidth]{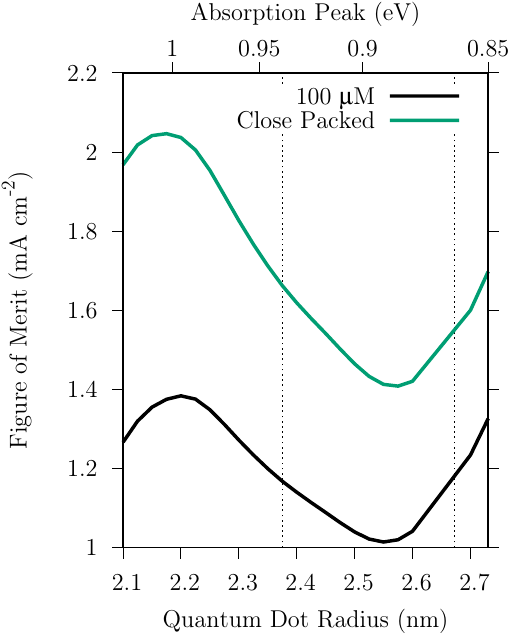}
\caption{Calculated fusion figure of merit of a \ce{PbS} quantum dot sensitized silicon solar cell.  The black curve assumes the sensitizer concentration is 100 $\mu$M.  The green curve assumes the sensitizer is close packed.  The anabathmophore thickness is optimized as previously reported \cite{frazer2017optimizing} with range \SIrange{1}{270}{\micro\meter}.  The dashed lines indicate a diameter change of one lattice parameter.}
\label{fig:fomsi}
\end{figure}
As shown in Fig. \ref{fig:fomsi}, we find that for silicon solar cells, the locally optimal \ce{PbS} quantum dot radius is around \SI{2.2}{\nm}, and the diameter is about seven lattice parameters.  Precision synthesis has been demonstrated on the scale of \SI{0.01}{\electronvolt}.\cite{cademartiri2006size} Quantum dots with a lower energy absorption peak (\SIrange{0.95}{0.85}{\electronvolt}, Supplementary Fig. 6) are less effective because of the lack of solar spectral irradiance at \SI{0.9}{\electronvolt} and increased self-absorption.  Smaller quantum dots are ineffective because of the reduced solar spectral irradiance at \SI{1.1}{\electronvolt}, which is just below the bandgap of silicon.  If the sensitizer has no absorption below the bandgap of silicon, the figure of merit is zero.  

A naive model of quantum dot performance would assume that a larger quantum dot with less quantum confinement implies a larger region of the solar spectrum will be captured. In addition, a larger quantum dot has a higher density of states and therefore absorbs more light at a given wavelength.  Owing to the nonlinearity in the exciton fusion rate equation, the naive model concludes that larger quantum dots always produce a device with a higher energy efficiency.  Our simulations show quantitatively that this naive model is conceptually correct, but omits four important details.  The first three are addressed here; the fourth was covered in our previous paper.\cite{frazer2017optimizing}

First, as shown in Fig. \ref{fig:absorption}, the near infrared spectrum of sunlight at the Earth's surface has a complicated structure, which is mostly influenced by Planck's law and the water in the atmosphere.  When the absorption peaks of a quantum dot sensitizer are tuned through the spectral irradiance peaks, the device efficiency is modulated.  

Second, excessively large quantum dots are disadvantageous owing to lower colloidal stability.\cite{shaw2013introduction,kister2018colloidal}  They have more traps per particle,\cite{chung2015size} reducing fusion efficiency.  Quantum dots must not overlap, indicating that concentration (number density) must eventually be sacrificed for large size.

Third, increased sunlight absorption by the sensitizer implies increased parasitic absorption of the upconverted light by the sensitizer.  Unlike some molecular sensitizers, quantum dots are never transparent at energies above the first excited state.

Fourth, to achieve efficient fusion, energy must be exothermically transferred from the sensitizer to the emitter molecule.  A larger quantum dot absorbs lower energy photons by having a lower exciton energy level. The exciton energy level limits the triplet energy level of the emitter molecule.\cite{frazer2017optimizing}  This level must be more than half the solar cell bandgap.  Other possible practical constraints on the triplet energy level merit investigation.

In Fig. \ref{fig:fomsi}, we find that a wide range of quantum dot radii and concentrations will produce a figure of merit exceeding \SI{1}{\milli\ampere\per\cm\squared}, which, if realized, would be an order of magnitude improvement on the state of the art,\cite{dilbeck2018molecular} despite the additional challenge of silicon compatibility.  The black curve is the results under the assumption that the quantum dot concentration is \SI{100}{\micro\textsc{m}}.  This is a reasonable concentration for quantum dots in solution.  In the absence of self-quenching,\cite{gholizadeh2018photochemical,jefferies2019photochemical} the figure of merit increases as a function of quantum dot concentration, as shown in Supplementary Fig. 5.

The green curve (Fig. \ref{fig:fomsi}) indicates figure of merit at the number density limit corresponding to close-packed (Kepler) quantum dots. The number density decreases with quantum dot size.  As a result, a small penalty is imposed on the performance of devices with larger quantum dots.  This effect shifts the optimal diameter by about one tenth of the \SI{5.9}{\angstrom} lattice parameter.  For a device with the maximum possible number density, a quantum dot which is too large can decrease the figure of merit by 31\%.

\begin{figure}[h]
\centering
\includegraphics[width=\linewidth]{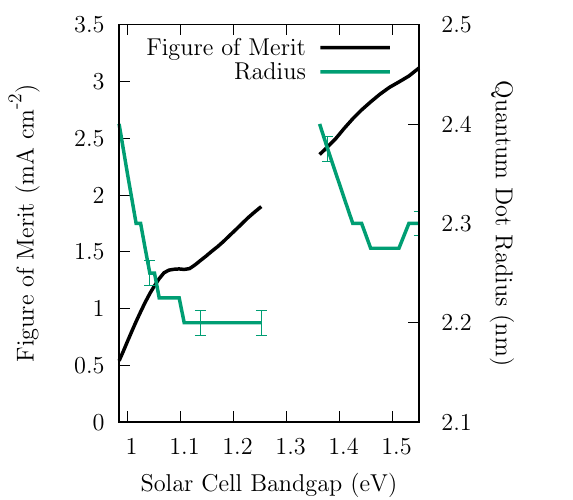}
\caption{Calculated optimal figure of merit and quantum dot size as a function of solar cell bandgap.  The anabathmophore thickness is optimized as previously reported \cite{frazer2017optimizing} with range \SIrange{66}{340}{\micro\meter}. Selected numerical error estimates for the quantum dot radius are shown as vertical bars.  No data is shown for solar cell bandgaps where larger quantum dots are always better.}
\label{fig:bandgap}
\end{figure}

Fig. \ref{fig:bandgap} shows the value of the optimal quantum dot radius and corresponding figure of merit as a function of solar cell bandgap.  Many laboratory solar cells have wider bandgaps than silicon (\SI{1.1}{\electronvolt}).  The Shockley-Queisser limit on solar cell efficiency is for an optimal bandgap of \SI{1.33}{\electronvolt}, but the upconversion detailed balance limit is at a bandgap of \SI{1.53}{\electronvolt}.\cite{jefferies2019photochemical}  While a bandgap exceeding \SI{1.53}{\electronvolt} will increase the figure of merit, it is unlikely to increase overall device performance.

The optimal quantum dot radius is discontinuous as a function of solar cell bandgap.  For smaller bandgaps, the excitonic absorption is tuned to match the first region of high spectral irradiance beneath the bandgap.  For larger bandgaps, multiple quantum dot absorption peaks are tuned to match multiple areas of high spectral irradiance.  For bandgaps near \SI{1.3}{\electronvolt}, the figure of merit increases monotonically as a function of quantum dot size.  In this situation, the best quantum dot cannot be selected without detailed knowledge of the corresponding emitter molecule and quantum dot colloidal stability.  The increase in figure of merit with bandgap found in Fig. \ref{fig:bandgap} by definition does not include the  benefit of the increase in cell voltage.

\section{Conclusions}
We conclude that \ce{PbS} quantum dot sensitizers have the absorption properties required to create upconversion devices with a figure of merit that far exceeds the current record,\cite{dilbeck2018molecular} even if constrained to be silicon-compatible.  While the total figure of merit will remain modest, \ce{PbS} sensitization is compatible in principle with all photovoltaic devices that have a bandgap bigger than bulk \ce{PbS}, which includes all plausible solar cell technologies.  Since light fusion technology can be synthesized in the liquid phase, it has the potential to be cheap enough that small figures of merit may be cost effective.  Under our assumptions, \ce{PbS} sensitization will exceed the performance required for economic benefit, which is estimated to be \SI{0.1}{\milli\ampere\per\square\cm}.\cite{peters2019value}

Our algorithm, which is distributed freely, can simulate scenarios with arbitrary illumination, absorption, and emission spectra.\cite{jefferies2019photochemical}  In this work we have addressed bandgaps and rate constants over the ranges relevant to photovoltaics.  Any quantum dot which possesses an absorption peak will sensitize most effectively when the absorption peak is at the same energy as an illumination peak.

To create a commercial device, it is necessary to create an emitter with suitable properties.  These include statistically and kinetically efficient energy transfer from the quantum dot sensitizer to the emitter, stable energy storage, rapid exciton transport, efficient annihilation, and efficient fluorescence.  All these properties have been demonstrated separately in experiments, but not simultaneously.

\begin{figure}[htbp]
\centering
\fbox{\includegraphics[width=.7\linewidth]{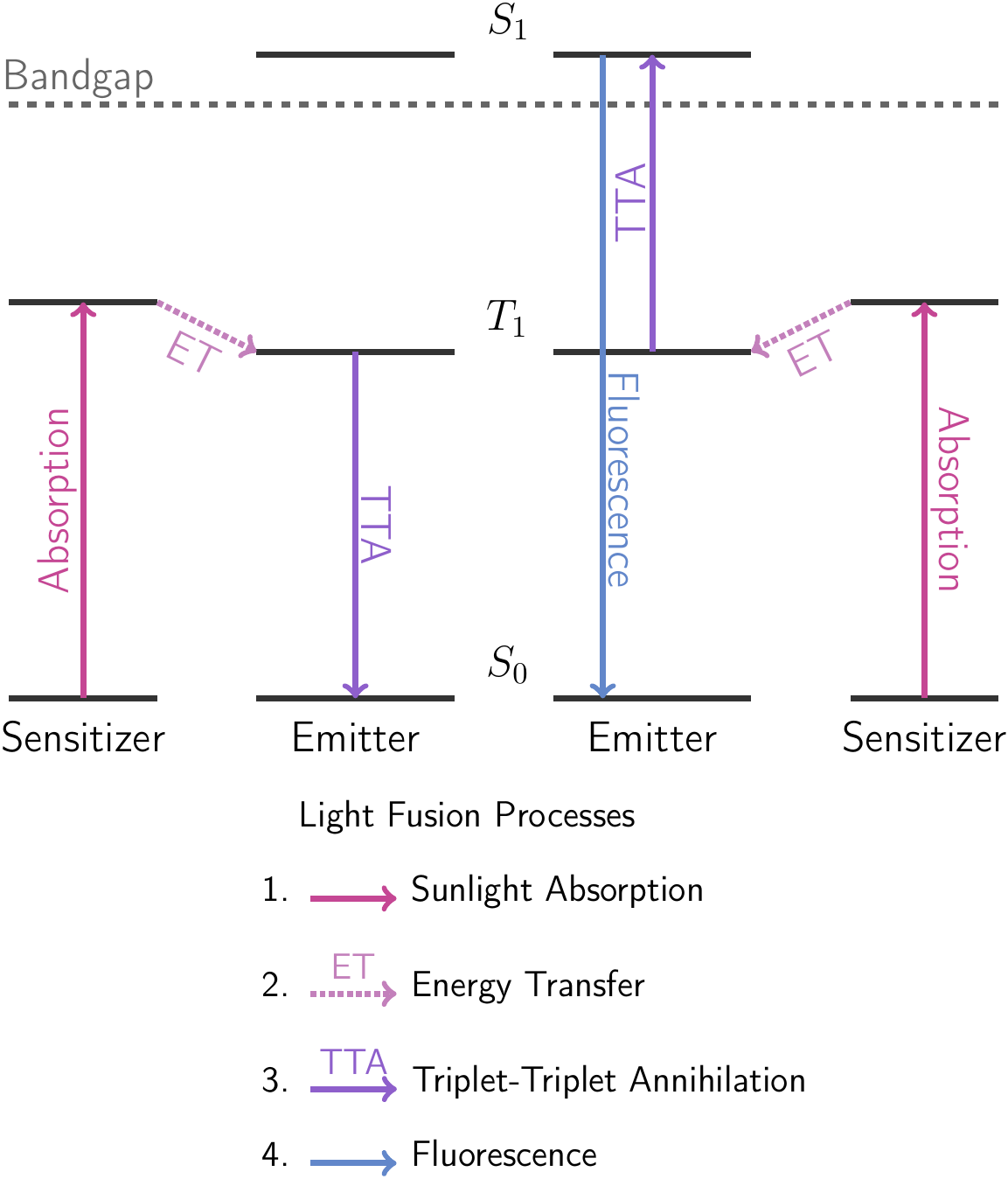}}
\caption{Energy level diagram for light fusion.  To be efficient, each process must be exothermic \cite{jefferies2019photochemical}. $S_n$ indicates the emitter molecule's $n$th singlet state.  $T_1$ indicates the first triplet state.}
\label{fig:el}
\end{figure}

\clearpage
\section*{Conflicts of interest}
There are no conflicts to declare.
\section*{Acknowledgements}
This research was undertaken with the assistance of resources and services from the National Computational Infrastructure (NCI), which is supported by the Australian Government.  B.S. acknowledges a Monash University ResearchFirst Fellowship.  We gratefully acknowledge financial support from from the Australian Research Council (CE170100026) via the ARC Center of Excellence in Exciton Science.  We gratefully acknowledge Prof. Ludovico Cademartiri and colleagues for providing the molar absorption data.  We acknowledge Elham M. Gholizadeh and Timothy W. Schmidt for helpful discussions. 

\section{Appendix: Supplementary Calculations}

\begin{figure}[htbp]
\centering
\fbox{\includegraphics[width=.5\linewidth]{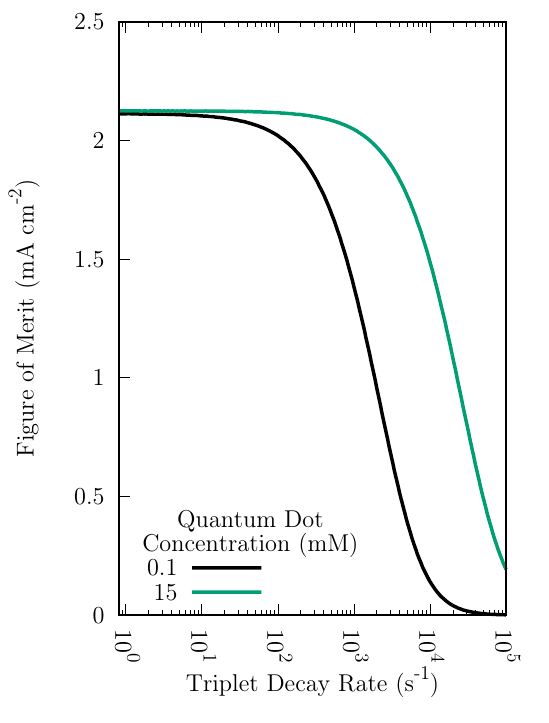}}
\caption{Figure of merit of quantum dot sensitized light fusion as a function of the decay rate for noninteracting triplet excitons. The decay rate is a property of the emitter \cite{gao2019intramolecular,pun2018tips}.  The decay rate can be as low as \SI{90}{\per \second} \cite{gholizadeh2018photochemical,gray2017loss,durandin2019critical}.  It can be increased by adding traps \cite{gholizadeh2018photochemical,jefferies2019photochemical} such as oxygen molecules.}
\label{fig:k1}
\end{figure}

\begin{figure}[htbp]
\centering
\fbox{\includegraphics[width=.5\linewidth]{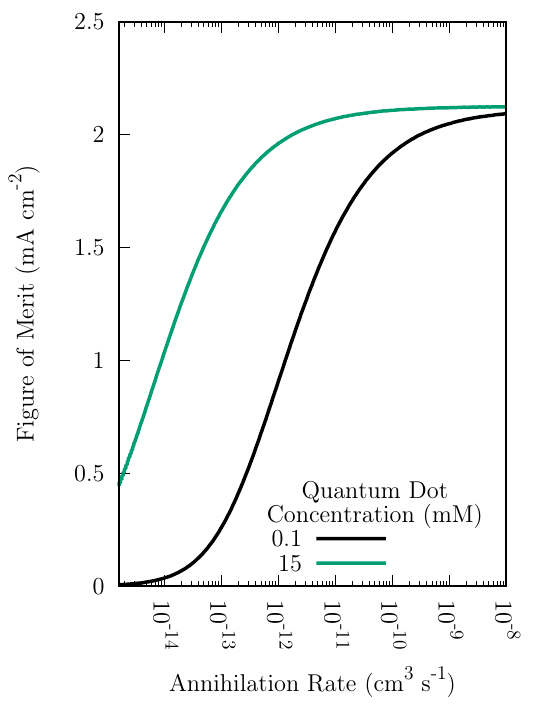}}
\caption{Figure of merit of quantum dot sensitized light fusion as a function of the annihilation rate constant for triplet excitons. The constant is a property of the emitter.  The annihilation rate constant varies from \SIrange{e-14}{e-9}{\cm\cubed\per\second} \cite{frazer2017optimizing,mahato2016preorganized,halas1997dynamics}.}
\label{fig:k2}
\end{figure}

\begin{figure}[htbp]
\centering
\fbox{\includegraphics[width=.5\linewidth]{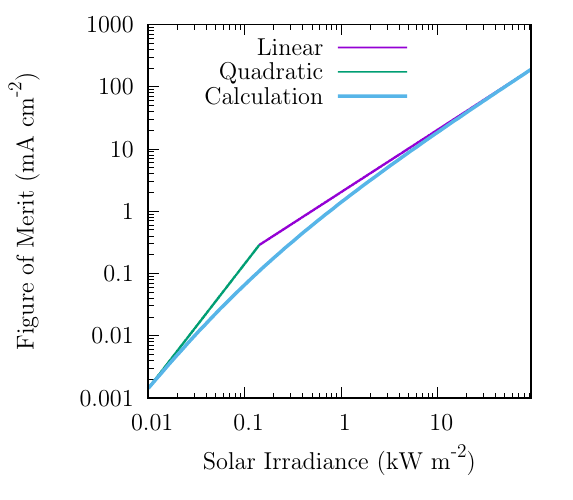}}
\caption{Figure of merit of quantum dot sensitized light fusion as a function of the solar irradiance.  The irradiance is measured before the solar cell, not at the sensitizer.  \SI{1}{\kilo\watt\per\square\meter} is the conventional value.  The transition from linear to quadratic behavior \cite{gray2018towards,yanai2016recent,haefele2012getting,gao2020triplet} happens near $I_\text{th}=$\SI{0.2}{\kilo\watt\per\square\m}.  Quadratic behavior indicates the quantum yield is at a maximum.  To the best of our knowledge, the lowest $I_\text{th}$ reported is \SI{0.09}{\kilo\watt\per\square\m} \cite{beery2019cdse}, but this is for monochromatic illumination in a silicon-incompatible device. }
\label{fig:sc}
\end{figure}

\begin{figure}[htbp]
\centering
\fbox{\includegraphics[width=.5\linewidth]{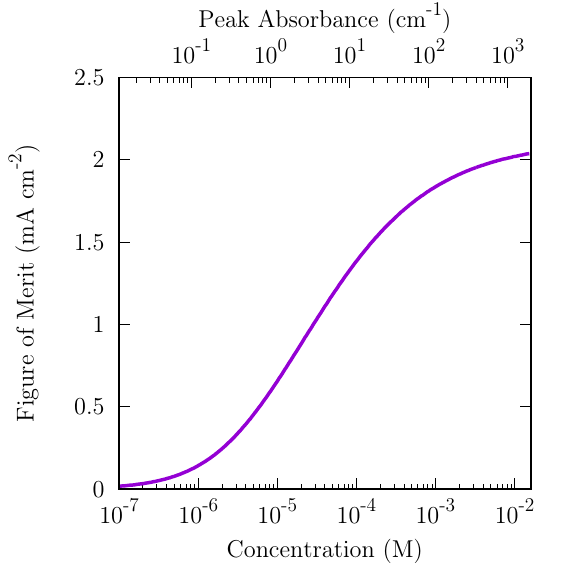}}
\caption{Figure of merit of quantum dot sensitized light fusion as a function of the quantum dot concentration \cite{pun2019annihilator,imperiale2019triplet}.  The absorbance of the quantum dot at the lowest energy absorption peak is also indicated.  A higher concentration results in a higher figure of merit.}
\label{fig:conc}
\end{figure}

The behavior of triplet fusion can be summarized by the well-known differential equation \cite{schmidt2014photochemical}:

\begin{align}
	\frac{d[T]}{dt}&=k_\phi [S]-k_1[T]-k_2[T]^2=0
\end{align}
Where $t$ is time, $[T]$ is the concentration of triplet excitons, $k_\phi$ is the sensitizer excitation rate, $[S]$ is the sensitizer concentration, $k_1$ is the decay rate for noninteracting triplet excitons, and $k_2$ is the annihilation rate constant for triplet excitons.  The $k_2[T]^2$ term produces the upconversion.  This equation qualitatively explains Figs. \ref{fig:k1}--\ref{fig:conc}.  Figures \ref{fig:k1} and \ref{fig:k2} show the calculated performance of quantum dot sensitizers when paired with emitters having various properties.  Fig. \ref{fig:sc} is calculated as a function of illumination conditions.  Fig. \ref{fig:conc} addresses quantum dot concentration.  Collectively, these figures show that a high ($>$\SI{1}{\milli\ampere\per\square\cm}) figure of merit can be achieved over a wide range of device types.  

\begin{figure}[htbp]
\centering
\fbox{\includegraphics[width=.5\linewidth]{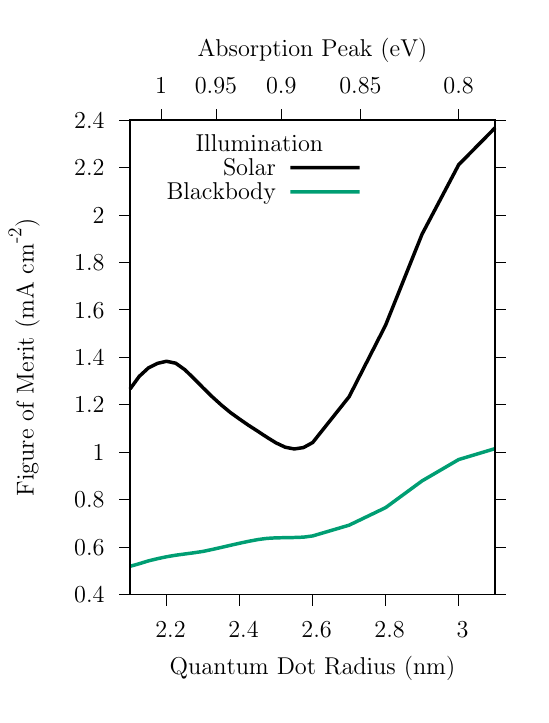}}
\caption{Figure of merit of quantum dot sensitized light fusion as a function of the quantum dot radius.  The black curve is the same as main text Fig. 3.  The green curve is the same calculation, except the solar spectrum is replaced by a \SI{5778}{\kelvin} blackbody spectrum \cite{stix2012sun}.  The irradiance is held constant at \SI{1}{\kilo\watt\per\square\cm}.  The attenuation by the earth's atmosphere is the main reason the solar spectrum is redder than a blackbody.   The blackbody approximation fails to capture the full importance of the quantum dot radius.}
\label{fig:bb}
\end{figure}

Fig. \ref{fig:bb} demonstrates that an insufficiently accurate model of the solar spectral irradiance should not be used to inform device design.

\begin{figure}[htbp]
\centering
\fbox{\includegraphics[width=.5\linewidth]{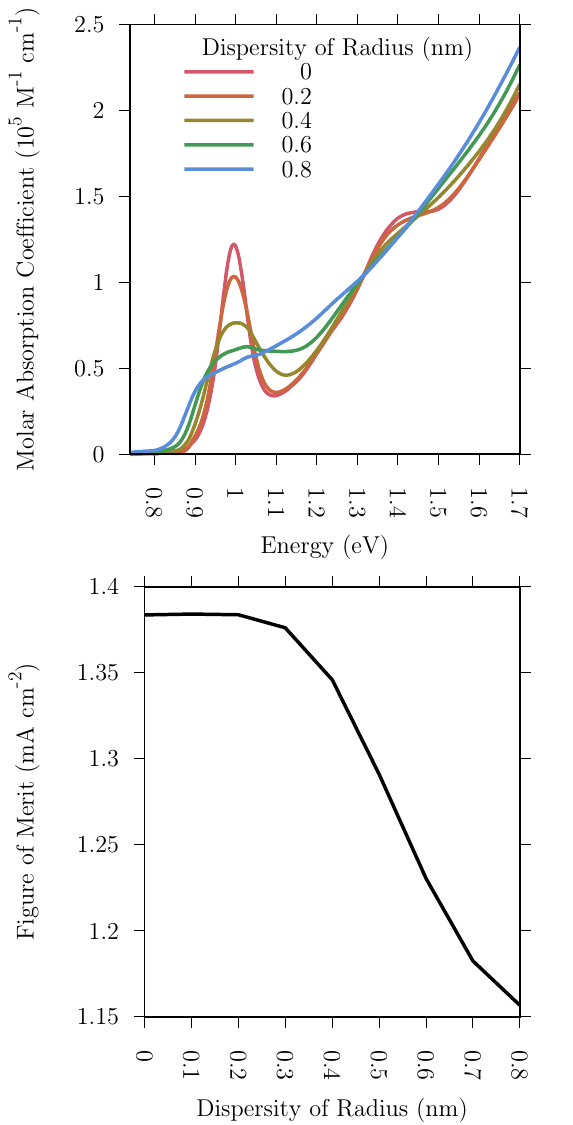}}
\caption{Molar absorption coefficient of quantum dots and figure of merit as a function of the dispersity of the quantum dot radius.}
\label{fig:dispersity}
\end{figure}

In our source for quantum dot molar absorptivity, the quantum dot dispersity is at most 10\% \cite{cademartiri2006size}.    Higher dispersity will reduce the degree to which the figure of merit depends on the quantum dot size.  For optimally sized quantum dots, dispersity slightly decreases the figure of merit slightly, as shown in Fig. \ref{fig:dispersity}. This is an example of regression towards the mean.   We define dispersity as the width of a rectangular distribution of radii centered at \SI{2.2}{\nano\meter}.  

In each supplementary figure, the thickness of the device is variable as a function of the horizontal axis.  The thickness was selected to produce the highest figure of merit \cite{frazer2017optimizing}.  The quantum dot radius is \SI{2.2}{\nano\meter}. The quantum dot concentration is  \SI{0.1}{\milli\textsc{M}}, except in Fig. \ref{fig:conc}.


\section{Appendix: Toxicity}

Toxicity is a concern for technologies that may be mass produced.  We estimate a lead sulfide density of \SI{6}{\gram\per\square\meter}.  The LC\textsubscript{50} toxicity metric of bulk lead sulfide in the fish \emph{Pimephales promelas} is \SI{0.9}{\milli\gram\per\liter} \cite{erten1998evaluation}.  We did not locate mammalian toxicity data or any data for \ce{PbSe}.  The ratio of density to LC\textsubscript{50} is \SI{7}{\meter} of water, which indicates that careful device disposal is required.  For \ce{PbS} nanomaterials, surface passivation may substantially reduce toxicity \cite{truong2011differential,liu2016non}.
\balance

\bibliography{bib} 
\bibliographystyle{rsc} 
\end{document}